\documentclass[runningheads]{svmult}
\usepackage{makeidx}   
\usepackage{graphicx}  
\usepackage{subeqnar}  
\usepackage{multicol}  
\usepackage{cropmark} 
\usepackage{physprbb}  
\newcommand{\be}[1]{\begin{equation} \label{(#1)}}
\newcommand{\ee}{\end{equation}}
\newcommand{\ba}[1]{\begin{eqnarray} \label{(#1)}}
\newcommand{\ea}{\end{eqnarray}}

\def\gsim{\mathrel{\vcenter{\hbox{$>$}\nointerlineskip\hbox{$\sim$}}}}

\begin{document}
\title*{Neutrinoless double beta decay potential \protect\newline 
in a large mixing angle world \thanks{Talk presented by H. P\"as at the
{\it DARK2000} Conference, Heidelberg, Germany.}}
%
%
\toctitle{The neutrino mass spectrum in a 
\protect\newline large mixing angle world}
%
%
\titlerunning{The neutrino mass spectrum}
%
\author{H.V. Klapdor--Kleingrothaus\inst{1}
\and H. P\"as\inst{2}
\and A. Yu. Smirnov\inst{3}
}
\authorrunning{Heinrich P\"as et al.}
%
%
\institute{Max--Planck--Institut f\"ur Kernphysik, P.O. Box 103980, 
D--69029 Heidelberg, Germany
\and 
Department of Physics and Astronomy, Vanderbilt University,
Nashville, TN 37235, USA
\and 
The Abdus Salam International Center of Theoretical Physics,
Strada Costiera 11, Trieste, Italy,
Institute for Nuclear Research, RAS, Moscow, Russia
}

\maketitle              

\begin{abstract}
We discuss the possibility of reconstructing the neutrino mass spectrum
from the complementary processes of neutrino oscillations and double beta 
decay in view of the new data of Super-Kamiokande presented at the
Neutrino2000 conference. Since the large mixing angle solution is favored,
now, the prospects to observe double beta decay and provide informations on
the absolute mass scale in the neutrino sector have been improved. 
\end{abstract}

\section{Double Beta decay and neutrino oscillations}

Neutrinos finally have been proven to be massive by atmospheric and 
solar neutrino oscillation experiments. However, the absolute scale of 
neutrino masses, a necessary ingredient for reconstructing beyond the standard
model physics, is still unknown, since informations obtained in
neutrino oscillation experiments regard the mass squared 
differences and mixing angles, only. Only {\it both} neutrino 
oscillations and neutrinoless double beta decay {\it together} 
could solve this absolute neutrino mass problem \cite{kps,prev,gluz}.
In this paper we discuss the most recent data
, as presented by the 
Super-Kamiokande Collaboration at the Neutrino2000 conference \cite{sk}.
The small mixing angle solution for solar neutrinos
is ruled out, now, at
90 \% C.L. 
Moreover, solutions including sterile neutrinos seemed to be
disfavored both for atmospheric as well as for solar neutrinos.
In the following we thus will restrict ourselves to a  three neutrino 
framework, omitting the LSND anomaly. (For a discussion of the small mixing
angle solution and four neutrino scena\-rios see \cite{kps}).
A global analysis in a three neutrino framework
yield the following favored regions 
\cite{gon99,hallo}:

\begin{itemize}

\item
Solar neutrino oscillations favor $\nu_e-\nu_{\not e}$ oscillations within
the large mixing angle (LMA) MSW solution:\\
$\Delta m_{\odot}^2= 3~(1-10)\cdot 10^{-5}$~eV$^2$\\
$\tan^2 \theta_{\odot}=0.5~(0.2-0.6)$,\\
where the bestfit is given with the 90 \% C.L. region in the brackets.\\
Also a small region in the QVO(quasi-vacuum-oscillation)-LOW regime at
$\Delta m_{\odot}^2= 10^{-7}$~eV$^2$,
$\tan^2 \theta_{\odot}=(0.6-0.8)$
is still allowed at 90 \% C.L., while disfavored compared to the small and 
large mixing solutions in an analysis of the neutrino energy spectra of
supernova 1987A \cite{SN87a}.

\item
Atmosheric neutrino oscillations are solved by $\nu_{\mu}-\nu_{\tau}$ 
oscillations with: \\
$\Delta m_{atm}^2 = 3~(1.6-5) \cdot 10^{-3}$~eV$^2$,\\
$\sin^2 2\theta_{atm} > 0.85$.

\end{itemize}

\noindent
Neutrinoless double beta ($0\nu\beta\beta$) decay
\be{}
^{A}_{Z}X \rightarrow ^A_{Z+2}X + 2 e^- 
\ee
has been shown to be a sensitive tool both for physics beyond the 
standard model \cite{cosmo99,pbsm} as well as for the
reconstruction of the neutrino mass spectrum \cite{kps}.
The most stringent limit is obtained from the Heidelberg--Moscow experiment
\cite{hdmo},
\be{}
\langle m \rangle = 0.27~{\rm eV}~~{\rm (68 \% C.L.)}.
\ee
Future experiments such as CUORE \cite{cuore}, MOON \cite{moon}
and EXO \cite{exo} and GENIUS \cite{gen} aim at sensitivities
down to $10^{-2}$ - $10^{-3}$ eV.

\begin{figure}[!t]
\begin{center}
\includegraphics[width=0.5\textwidth]{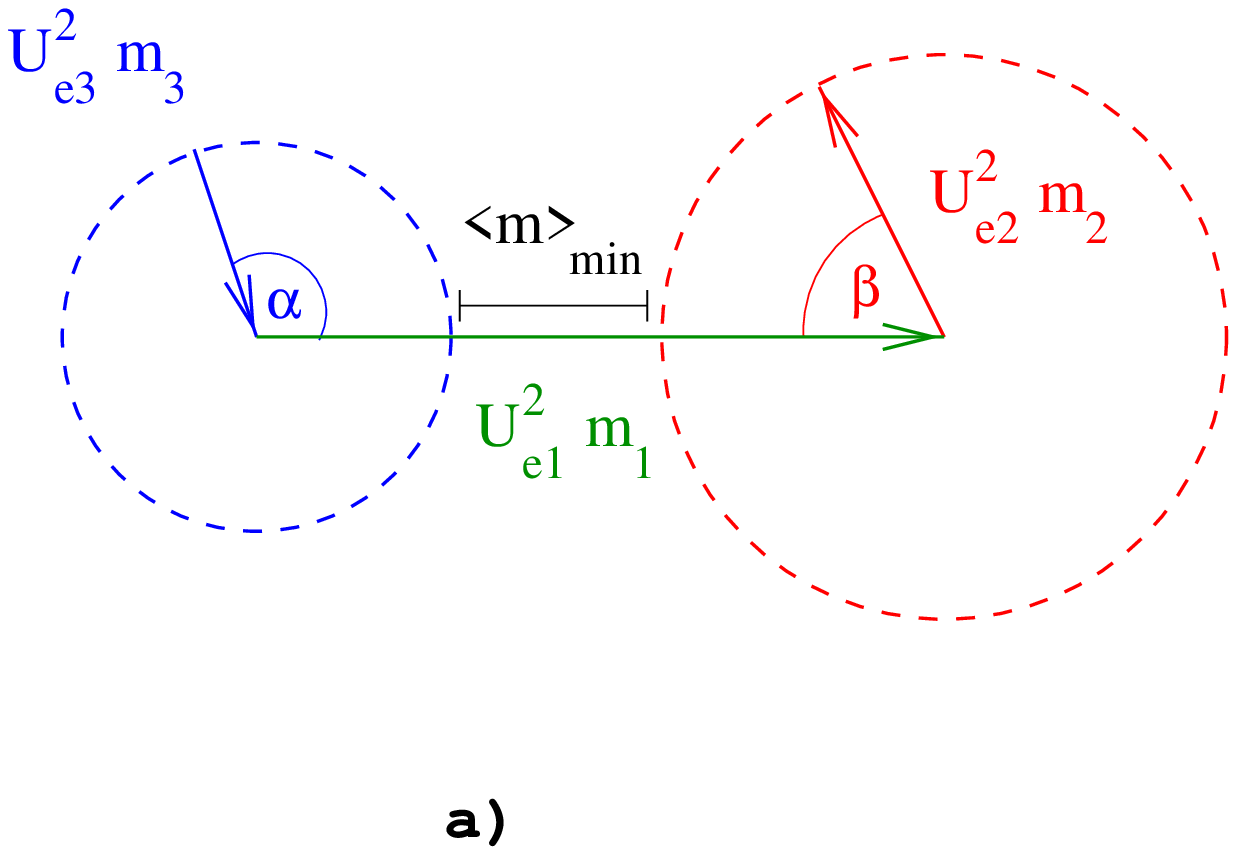}
\includegraphics[width=0.5\textwidth]{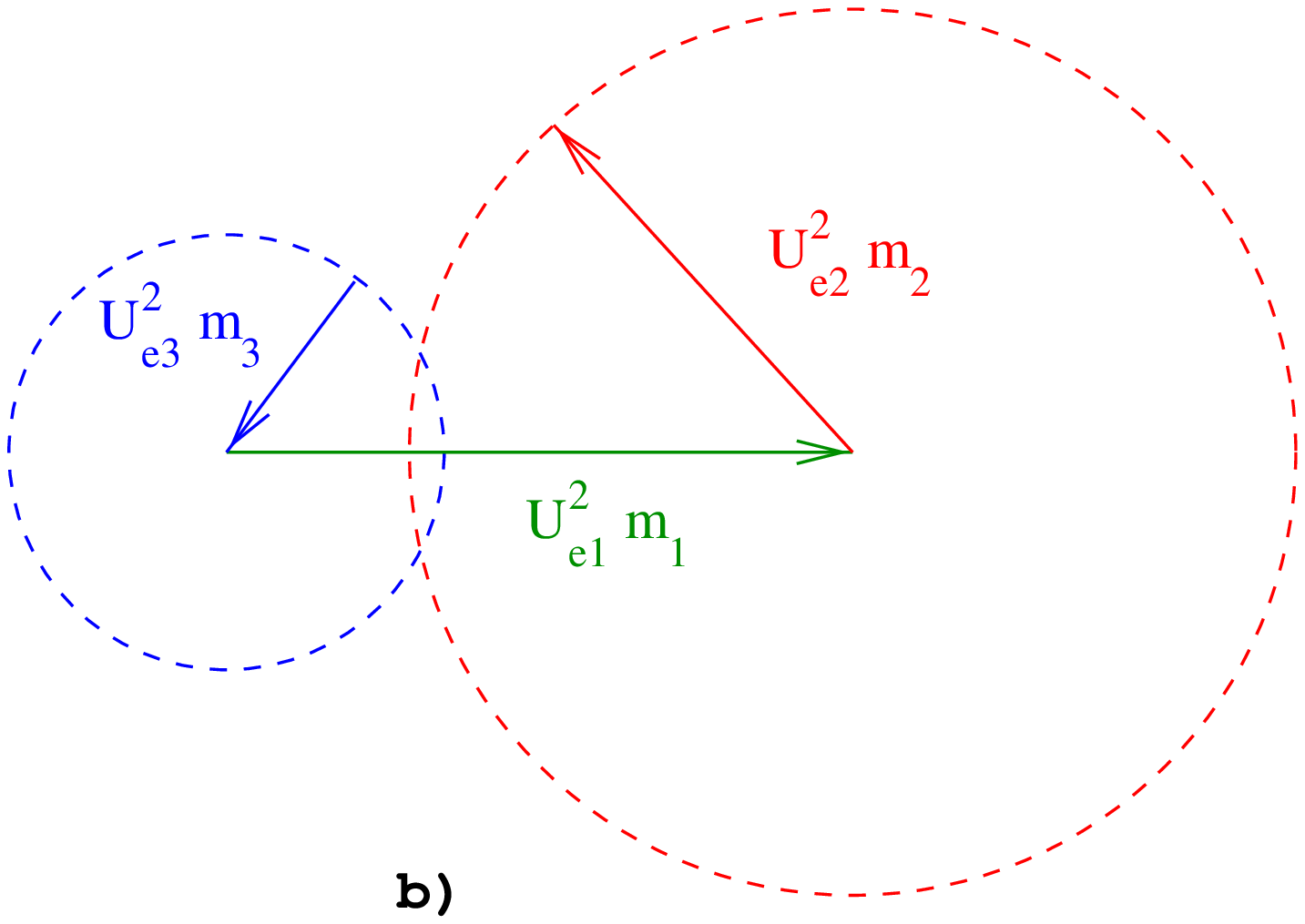}
\end{center}
\caption[]{The effective Majorana mass $\langle m \rangle$ in the 
complex plane. Vectors show contributions to $\langle m \rangle$ 
from individual eigenstates. 
The total $\langle m \rangle$  appears as the sum of the three vectors. 
Allowed values of $\langle m \rangle$ correspond to modulies of
vectors which connect two points on the circles. Here
$\alpha = \phi_{3}-\pi$,  $\beta= \pi-\phi_{2}$.
a). $|m^{(1)}_{ee}|>|m^{(2)}_{ee}|+|m^{(3)}_{ee}|$: the 
vectors $\vec{m}^{(i)}_{ee}$ can not form a triangle 
and no complete cancellation occurs. b)   
$|m^{(1)}_{ee}|\leq|m^{(2)}_{ee}|+|m^{(3)}_{ee}|$:   
in this case complete cancellation occurs in the intersection points of
the circles, so that $\langle m \rangle = 0$.
(from \protect{\cite{kps}}).
\label{graph2}}
\end{figure}

The observable measured in the mass mechanism of $0\nu\beta\beta$ decay
is the $ee$ element of the neutrino mass matrix in flavor space,
the effective neutrino mass
\be{}
\langle m \rangle = |\sum U_{ei}^2 m_i|,
\ee
where $U_{ei}$ denote the elements of the neutrino mixing matrix.
For the three-neutrino case we get 
\be{}
\langle m \rangle = 
|m^{(1)}_{ee}| + e^{i\phi_{2}} |m_{ee}^{(2)}|
+  e^{i\phi_{3}} |m_{ee}^{(3)}|~,
\label{mee}
\ee
where $m_{ee}^{(i)}\equiv |m_{ee}^{(i)}| \exp{(i \phi_i)}$ 
($i = 1, 2, 3$)  are  the contributions to $\langle m \rangle$
from individual mass eigenstates, which can be written in terms 
of oscillation parameters as:  
\ba{  }
|m^{(1)}_{ee}|&=&|U_{e1}|^2 m_1, \\
|m^{(2)}_{ee}|&=&|U_{e2}|^2 \sqrt{\Delta m^2_{21} + m_1^2},\\
|m^{(3)}_{ee}|&=&|U_{e3}|^2\sqrt{\Delta m^2_{32}+ \Delta m^2_{21} + m_1^2},
\label{gg}
\ea
and  $\phi_{i}$  are the relative Majorana CP-phases. 
The contributions $m_{ee}^{(i)} $ 
can be illustrated as vectors in the complex
plane (fig. \ref{graph2}).

Some of the parameters in eq. \ref{gg} can be fixed or restricted 
from neutrino oscillation data: In the case of normal hierarchy 
$\Delta m^2_{21}$, 
$|U_{e1}|^2=\cos^2 \theta_{\odot}$ and $|U_{e2}|^2=\sin^2 \theta_{\odot}$ 
can be obtained from solar neutrinos, 
$\Delta m^2_{32}$ from atmospheric neutrinos
and $|U_{e3}|^2$ is restricted from experiments searching for electron 
disappearance such as CHOOZ. For inverse hierarchy one has to exchange 
neutrinos $\nu_1 \leftrightarrow \nu_3$ in the equations. 
The phases $\phi_i$ and the mass of the lightest 
neutrino, $m_1$, are free parameters. Thus the search for neutrinoless double
beta decay can provide informations about the neutrino mass spectrum and 
the absolute mass scale.  
With increase of $m_1$ the level of degeneracy of the neutrino
spectrum increases and we can distinguish the extreme cases 
of hierarchical spectra,
$m_1^2 \ll \Delta m^2_{21} \ll  \Delta m^2_{31}$
and degenerate spectra
$\Delta m^2_{21} \ll  \Delta m^2_{31}  \ll m_1^2$.    
In the following we discuss these extreme cases as well as
transition regions in detail, and comment on the case of inverse hierarchy.

\section{Hierarchical spectra}
\begin{figure}[!th]
\begin{center}
\includegraphics[width=0.8\textwidth]{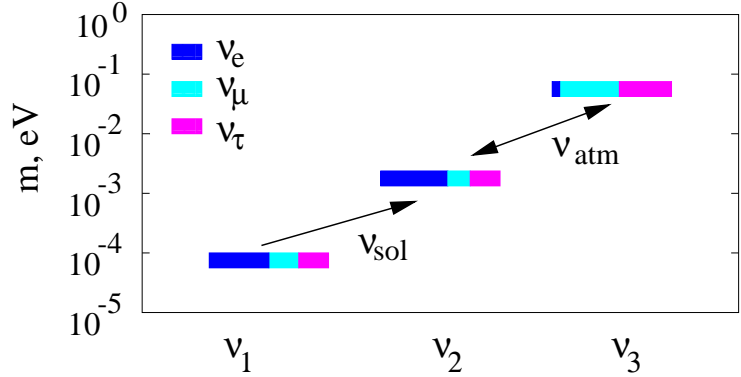}
\end{center}
\caption[]{Neutrino masses and mixings in the scheme with mass hierarchy.
Coloured bars correspond to flavor
admixtures in the mass eigenstates $\nu_1$, $\nu_2$, $\nu_3$. The quantity
$\langle m \rangle$ is determined by the dark blue bars denoting the admixture
of the electron neutrino $U_{ei}$.
\label{smi1}}
\end{figure}

Hierarchical spectra (fig. \ref{smi1})
\be{}
m_1 \ll m_2 \ll m_3
\ee
can be motivated by analogies with the quark sector and the simplest see-saw 
models. In these models the contribution of $m_1$ to the double beta decay 
obser\-vable $\langle m \rangle$
is small. The main 
contribution is obtained from $m_2$ or $m_3$, depending on the solution of the
solar neutrino deficit.

\begin{figure}[!ht]
\begin{center}
\includegraphics[width=1.0\textwidth]{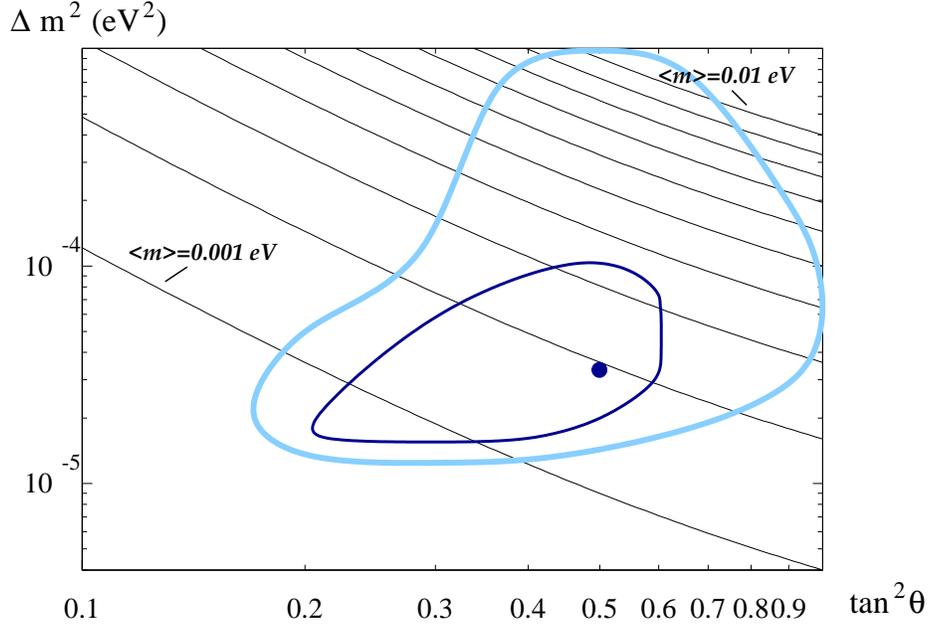}
\end{center}
\caption[]{
Double beta decay oberservable $\langle m \rangle$ and oscillation parameters:
The case for the MSW large mixing solution of the solar 
neutrino deficit, where the dominant contribution to $\langle m \rangle$
comes from the second state, shown are lines 
of constant $\langle m \rangle$.
The inner and outer closed line show the regions allowed by present 
solar neutrino experiments with 90 \% C.L. and 99 \% C.L., respectively. 
Complementary informations can be obtained
from double beta decay, the search for
a day-night effect and spectral distortions 
in future solar neutrino experiments as well as a disappearance signal 
in KAMLAND. 
\label{dark2}}
\end{figure}

After {\it Neutrino2000}, the prospects of a positive signal in double 
beta decay are more promising, now.
If the large mixing solution of the solar neutrino deficit 
is realized,
the contribution of $m_2$ becomes dominant due to the almost maximal 
$U_{e2}$ and the relatively large $\Delta m^2_{21}$:
\be{}
\langle m \rangle \simeq m_{ee}^{(2)}= 
\frac{\tan^2 \theta}{1+ \tan^2 \theta} \sqrt{\Delta m_{\odot}^2}.
\ee

Fig. \ref{dark2} shows values of $\langle m \rangle$ 
in the range of
the large mixing angle solution. The closed lines denote the regions allowed
at 90 \% C.L. and 99 \% C.L. according to \cite{gon99}.
In the 90 \% C.L. region the prediction for $\langle m \rangle$ becomes
definite, now, 
\be{}
\langle m \rangle = (1-3) \cdot 10^{-3}~{\rm eV}.
\ee
A coincident measurement 
of $\langle m \rangle$ at this order of magnitude with corresponding results 
of day-night asymmetry and energy spectra of solar neutrino rates 
together with a confirmation of the large mixing angle solution 
by the long baseline reactor experiment KAMLAND \cite{kam}
would identify a single point in the large mixing angle MSW 
solution and provide a strong hint for this scheme. 

It should be stressed, that a large portion of the 99 \% C.L. favored 
region extends to large $\Delta m_{\odot}^2$ 
allowing for effective neutrino Majorana
masses well above $10^{-2}$ eV even in the hierarchical case.

\begin{figure}[!ht]
\begin{center}
\includegraphics[width=1.0\textwidth]{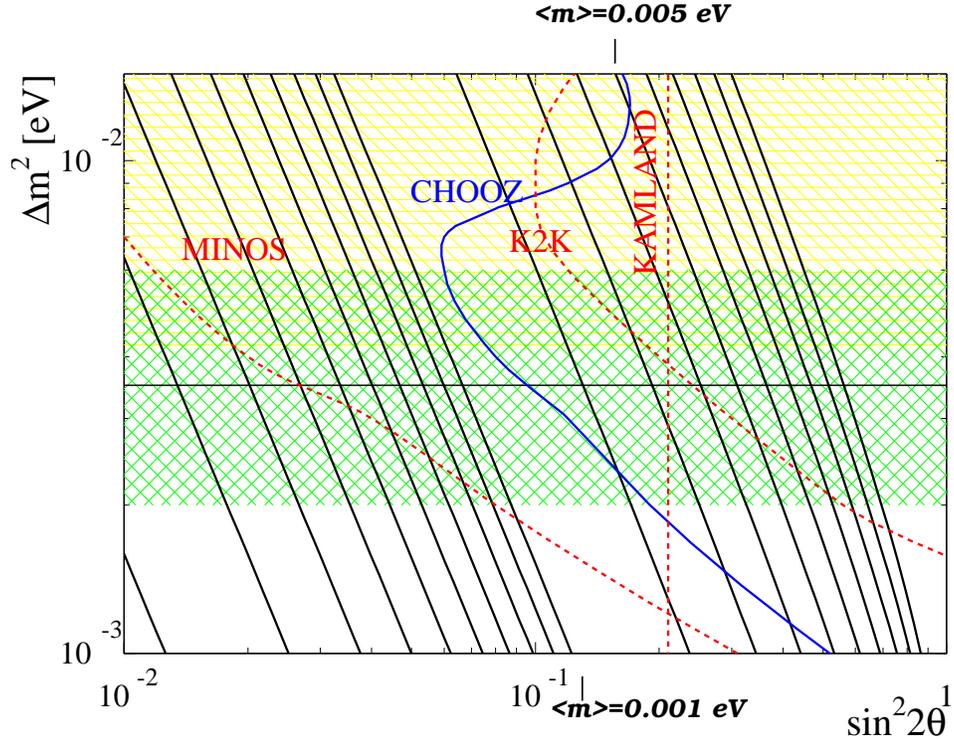}
\end{center}
\caption[]{
Double beta decay observable $\langle m \rangle$ and oscillation parameters:
The case of hierarchical schemes with the QVO-LOW solution. Shown is the dominant
contribution of the third state to $\langle m \rangle$
which is constrained by the
CHOOZ experiment, excluding the region to the upper right. Further 
informations can be obtained from the long baseline project MINOS
and future double beta decay experiments \protect{\cite{kps}}.
\label{dark1}}
\end{figure}

If the less favored QVO-LOW solution is realized in solar neutrinos,
$U_{e2}$ is close to maximal but the 
mass of the second state is tiny. In these cases the main contribution 
to $\langle m \rangle$ comes 
from $m_3$:
\be{}
\langle m \rangle \simeq m_{ee}^{(3)} = \frac{1}{4} \sqrt{\Delta m_{atm}^2}
\sin^2 2 \theta_{ee},
\ee 
where $\sin^2 2 \theta_{ee}=4 U_{e3}^2$ 
denotes the mixing angle restricted in
disappearance experiments.
The situation is illustrated in fig. \ref{dark1}. 
Here lines of constant $\langle m \rangle$
are shown as functions of the oscillation parameters $\Delta m_{13}^2$
and $\sin^2 2 \theta_{ee}$. 
The shaded areas show the mass $m_3 \simeq \sqrt{\Delta m^2_{13}}$
favored by atmospheric neutrinos with the horizontal line indicating 
the best fit value. The region to the upper right is excluded by the 
nuclear reactor 
experiment CHOOZ \cite{chooz}, 
implying $\langle m \rangle<2 \cdot 10^{-3}$ eV in the range favored by
atmospheric neutrinos. Obviously 
in this case only the 10 ton GENIUS experiment could observe a positive 
$0\nu\beta\beta$ decay signal. A coincidence of such a measurement with 
a oscillation signal at MINOS  
and a confirmation of the solar QVO-LOW MSW oscillations 
by solar neutrino experiments would 
be a strong hint for this scheme.

\section{Degenerate Scenarios}
\begin{figure}[!t]
\begin{center}
\includegraphics[width=0.8\textwidth]{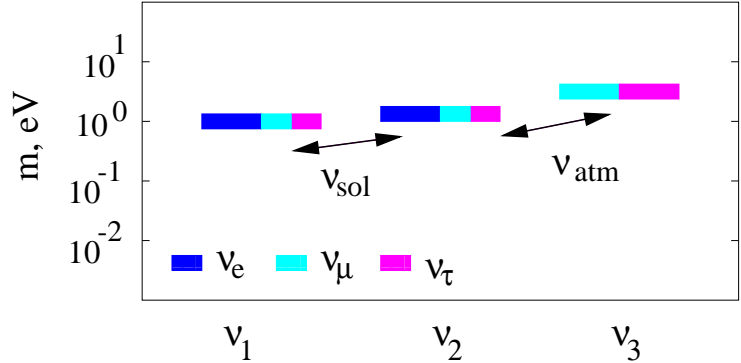}
\end{center}
\caption{Neutrino masses and mixings in the degenerate scheme.
\label{smi2}}
\end{figure}

In degenerate schemes (fig. \ref{smi2})
\be{}
m_1 \simeq m_2 \simeq m_3 \gsim 0.1 {\rm eV}
\ee
neutrinos still may be of cosmological relevance. 
Neutrinos with an overall mass scale of a few eV could play an 
important role as ``hot dark matter'' component of the universe.
When structures were formed 
in the early universe, overdense regions of (cold) dark matter provide the 
seeds of
the large scale structure, which later formed galaxies and 
clusters. A small
``hot'' (relativistic) component could prevent an overproduction of 
structure at small scales. 
Since structures redshift photons, this should imply also imprints 
on the cosmic microwave background (CMB), 
which could be measured by the future 
satellite experiments MAP and Planck \cite{cmb}. 
While this option of cold-hot-dark-matter cosmology has 
been disfavored by models including a cosmological constant, as supported
by the supernova cosmology project, 
a new motivation for degenerate models with a less large mass 
scale may come from the Z-burst interpretation of ultra high energy 
cosmic rays (UHECRs). In this model UHECRs are understood as the decay 
products of a resonant annihilation process of high energetic neutrinos
with the relic neutrino background \cite{weil}. Since the neutrino mass 
scale is related to the UHECR energy and relic neutrino clustering on 
galactic scales may turn out to be a necessary ingredient of the model,
an absolute neutrinos mass scale of $\sim 0.1-1$ eV
is predicted in this context
\cite{weil,wp}.

In degenerate schemes the mass differences are not significant. Since 
the contribution of $m_3$ is strongly bounded by CHOOZ again,
the main contributions to $\langle m \rangle$ come from $m_1$ and $m_{2}$,
which may cancel as an effect of the unknown Majorana CP-phases.
The relative contributions of these states
depend on their admixture of the electron flavor, which is determined by the
solution of the solar neutrino deficit.
Then the effective neutrino mass becomes 
\be{}
m_{min} < \langle m \rangle < m_1
\ee 
with
\ba{}
\langle m \rangle_{min} &=& (\cos^2 \theta_{\odot}-\sin^2 
\theta_{\odot})~m_1 \nonumber \\
      &=& \frac{1-\tan^2 \theta_{\odot}}{1+\tan^2 \theta_{\odot}}~m_1. 
\ea
This implies
\be{}
\langle m \rangle = (0.25 - 1) \cdot m_1
\ee
for the large mixing angle solution and
\be{}
\langle m \rangle = (0.1 - 1) \cdot m_1
\ee
for the QVO-LOW solution, where the range allowed corresponds to possible 
values of the unknown Majorana CP-phases. 
It should be stressed that this way an upper bound on the mass scale of the
heaviest neutrino can be deduced from the recent limit on $\langle m \rangle$.
For the LMA solution we obtain $m_{1,2,3}< 1$ eV, implying 
$\sum_i m_{i} < 3 eV$. For the QVO-LOW solution we obtain $m_{1,2,3}< 3$ eV, 
implying $\sum_i m_{i} < 9$ eV.

\begin{figure}[!ht]
\begin{center}
\includegraphics[width=1.0\textwidth]{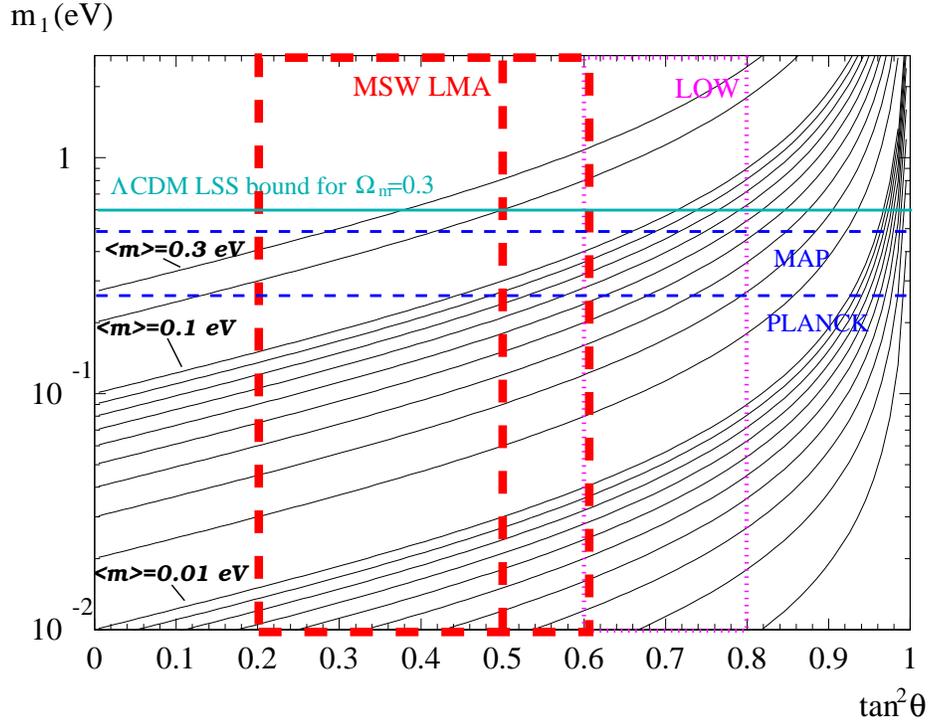}
\end{center}
\caption[]{Double beta decay oberservable $\langle m \rangle$ and 
oscillation parameters:
The case for degenerate neutrinos. Plotted on the axes are 
the overall scale of neutrino masses $m_0$
and the mixing $\tan^2 \theta_{\odot}$. 
The dashed boxes indicate the 90 \% C.L. allowed regions for the
large mixing angle (thick dashes, bestfit indicated also)
and LOW-QVO solution (thin dashes). 
Allowed values for $\langle m \rangle$ for a given $m_0$ correspond to the 
regions 
between $m_0$ and the corresponding curved line. Also shown 
is a cosmological bound obtained from a fit to the CMB and 
large scale structure and the expected
sensitivity of the satellite experiments MAP and Planck \protect{\cite{cmb}}. 
\label{dark3}}
\end{figure}

In fig. \ref{dark3} lines of constant double beta decay observables 
(solid curved lines) as functions of the solar mixing
are shown together with information from 
cosmological observations about the overall mass scale (horizontal lines). 
Shown is the bound $m_j < 0.6$~eV for each of three degenerate neutrinos
and for $\Omega_m=0.3$ at 95 \% C.L., obtained from a combined
fit to the CMB and large scale structure (LSS) data
(The constraint becomes $\sum_j m_j<5.5$~eV for arbitrary values 
of $\Omega_m$).
Also shown are the expected sensitivities of MAP and Planck 
to a single
neutrino state, 0.5 eV and 0.25 eV, respectively, including polarization 
data \cite{cmb}.

A coincidence of the absolute mass scale reconstructed from double beta decay 
and neutrino oscillations with a direct measurement of the neutrino mass
in tritium beta decay 
spectra \cite{trit} or its derivation from cosmological parameters 
determined from the CMB in the satellite experiments MAP and Planck 
and future LSS surveys
would 
prove this scheme to be realized in nature. To establish this triple evidence 
however is difficult due to the 
restricted sensitivity of the latter approaches. Future tritium experiments
aim at a sensitivity down to $\cal{O}$(0.1~eV) and MAP and Planck have been
estimated to be sensitive to $\sum m_{\nu}=0.5-0.25$ eV.
Thus for neutrino mass scales below $m_0 < 0.1$ eV only a range for the 
absolute mass scale can be fixed by solar neutrino experiments and double beta 
decay. 

The same conclusions are true for partially degenerate schemes, 
\be{}
m_1 \simeq m_2 \ll m_3,
\ee
keeping in 
mind that in these cases only the heaviest neutrino affects cosmology. 
The mass range
for partial degeneracy is $m_1 \sim 0.01 - 0.1$ eV
 
\section{Inverse Hierarchy}

\begin{figure}[!t]
\begin{center}
\includegraphics[width=0.8\textwidth]{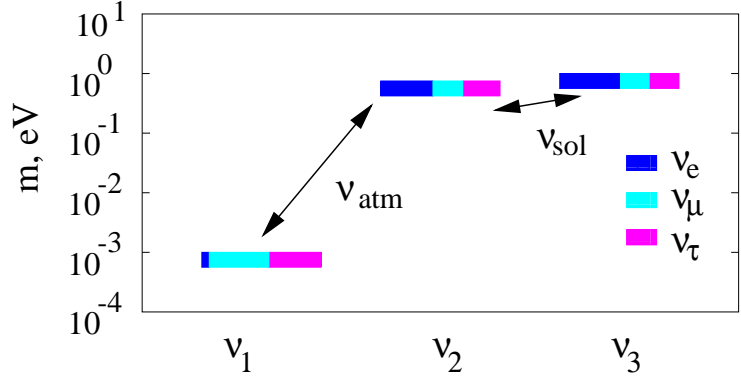}
\end{center}
\caption{Neutrino masses and mixings in the scheme with inverse hierarchy. 
\label{smi3}}
\end{figure}

A further possibility is an inverse hierarchical spectrum (fig. \ref{smi3})
\be{}
m_3 \simeq m_2 \gg m_1
\ee
where the heaviest state with mass $m_3$
is mainly the electron neutrino, now. 

Its mass is 
determined by the atmospheric neutrinos, $m_3 \simeq \sqrt{\Delta m^2_{atm}}$,
implying
\be{}
 \sqrt{\Delta m_{atm}^2} \frac{1-\tan^2 \theta_{\odot}}
{1+ \tan^2 \theta_{\odot}} < \langle m \rangle < \sqrt{\Delta m_{atm}^2}.
\ee
For both the large mixing MSW or QVO-LOW 
solution cancellations of the two heavy states become possible
and $\langle m \rangle=(1-7) \cdot 10^{-2}$ eV,
$\langle m \rangle=(0.4-7) \cdot 10^{-2}$ eV, respectively. 
A test of the inverse hierarchy 
is possible 
in matter effects of neutrino oscillations. For this case the MSW
level crossing 
happens for antiparticles rather than for particles. Effects could be 
observable in long baseline experiments and in the neutrino spectra of 
supernovae \cite{supn}. In fact a recent analysis \cite{inv}
of SN1987A obtains a strong indication that the inverted mass 
hierarchy is disfavored unless $U_{e1}$ is large.

\section{Transition Regions}

\begin{figure}[!ht]
\begin{center}
\includegraphics[width=0.9\textwidth]{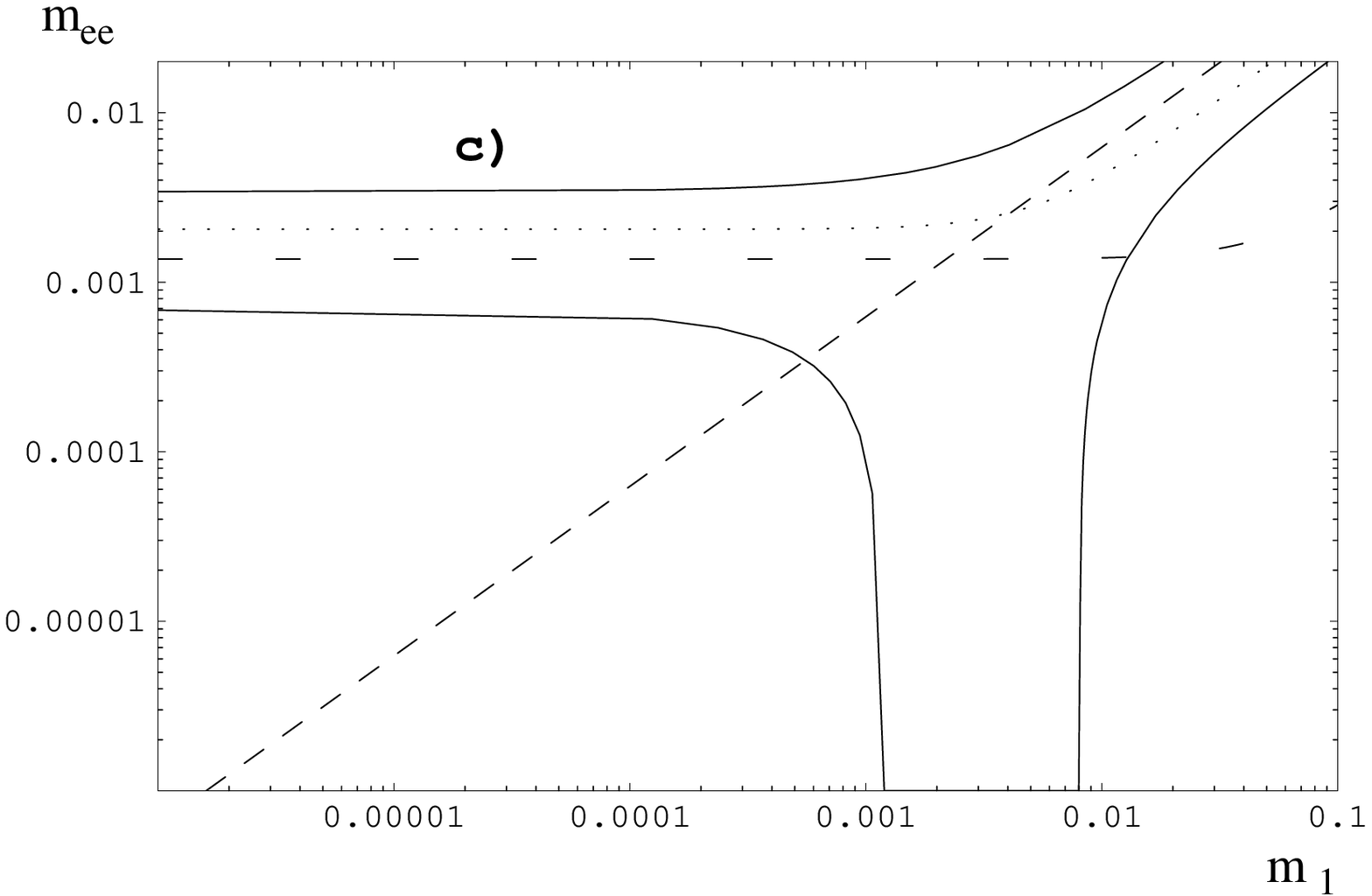}
\end{center}
\caption[]{$\langle m \rangle$ 
(eV) as a function of $m_1$ (eV) for three-neutrino mixing.
Shown are the contributions $m_{ee}^{(1)}$ (dashed), $m_{ee}^{(2)}$
(dotted) and $m_{ee}^{(3)}$ (interrupted dashes).
The solid lines correspond to $\langle m \rangle_{max}$ and 
$\langle m \rangle_{min}$ and show 
the allowed region for $\langle m \rangle$.
Panels a)-c) correspond to the cases for $U_{e2}^2=0.17$, $U_{e2}^2=0.33$,
and $U_{e2}^2=0.38$, i.e. the small mixing bound, best fit and large mixing 
bound of
the 90 \% C.L. level LMA solution.
The mixing of
the third state is varied from zero to its upper bound,
$U_{e3}^2=2.5 \cdot 10^{-2}$.
\label{lmatrans}}
\end{figure}

In fig. \ref{lmatrans} we show the dependence of the individual
contributions $m_{ee}^{(i)}$ to $\langle m \rangle$ on $m_1$,
for different values of mixing within the LMA solution.
Panel a)-c) correspond to the small mixing bound, best fit and large mixing 
bound
of the 90 \% C.L. allowed region, respectively.   
For $m_{ee}^{(3)}$ only the upper bound is used; the two other 
lines represent possible values of $m_{ee}^{(1)}$ and $m_{ee}^{(2)}$ 
for the specific 
neutrino mixing parameters.  
We show also the maximal and the minimal possible values of 
$\langle m \rangle$.

\begin{figure}[!ht]
\begin{center}
\includegraphics[width=0.9\textwidth]{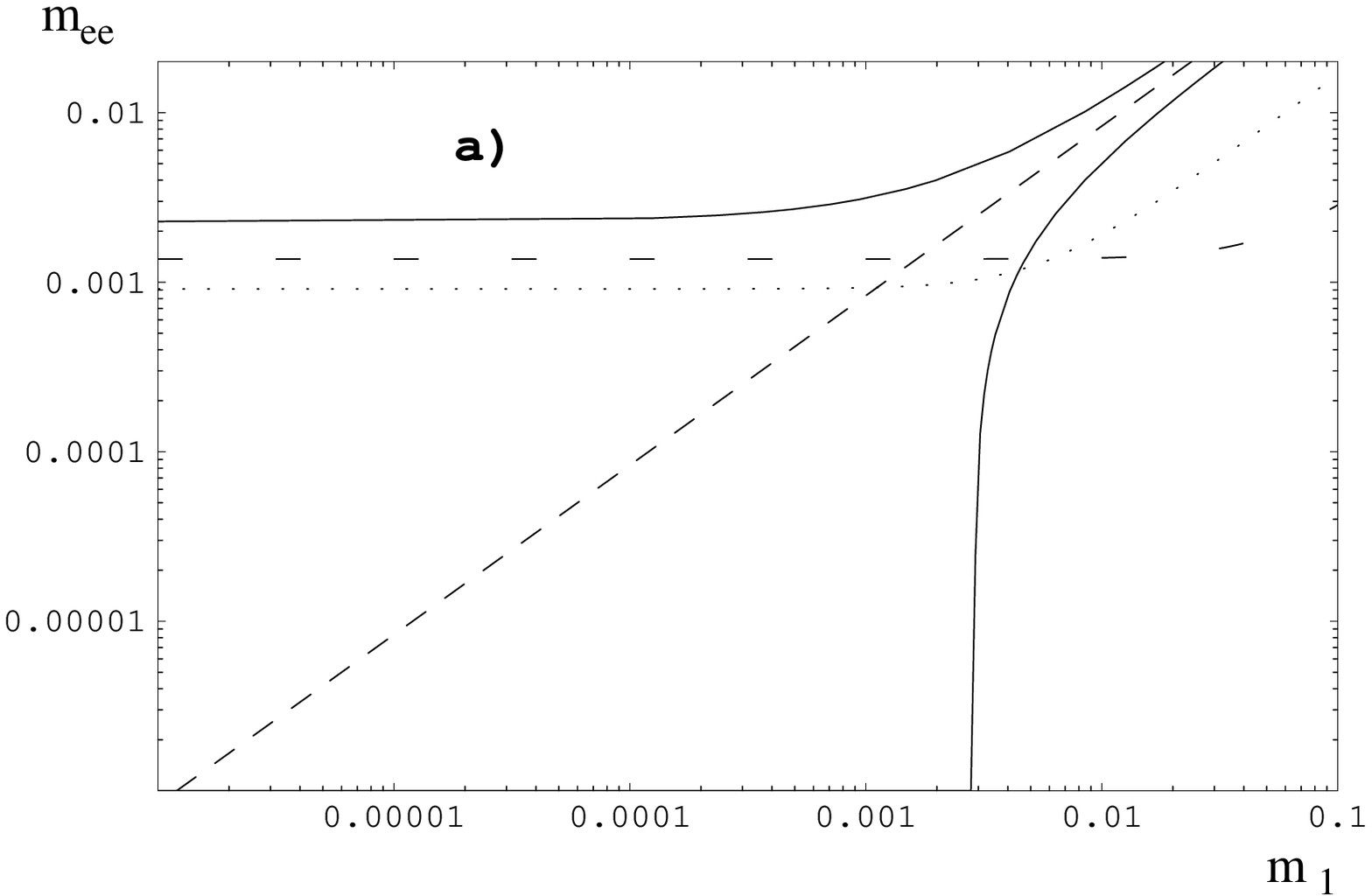}
\includegraphics[width=0.9\textwidth]{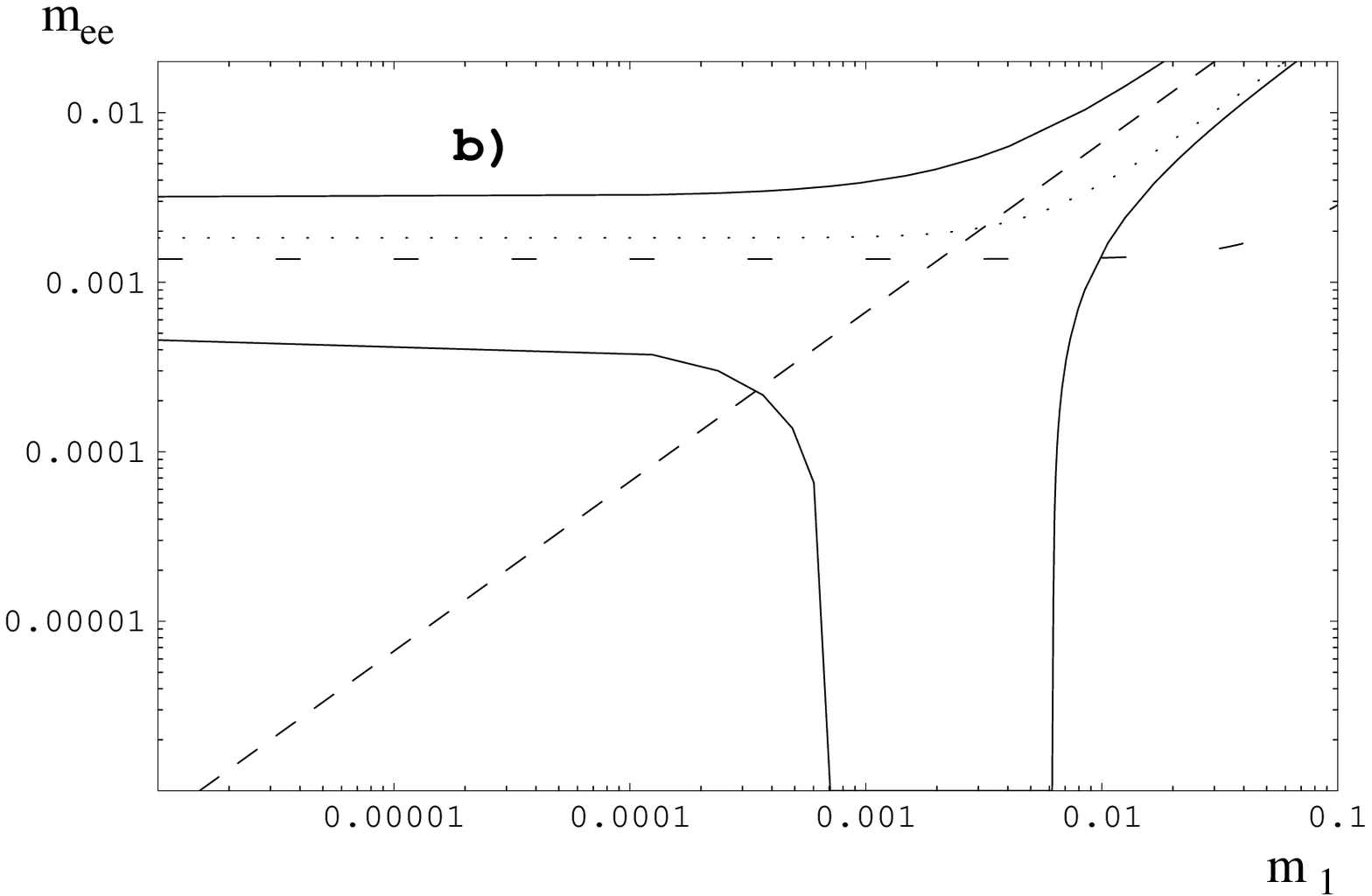}
\end{center}
\end{figure}

The upper bounds on $\langle m \rangle$  as functions of $m_1$ have a similar
dependence for all the cases. The lower bound in the hierarchical region
($m_1  < 10^{-3}-10^{-2}$~eV) crucially depends the solar mixing angle. 
If the solar mixing is sufficiently large
the contribution 
from $m_2$ dominates and no cancellation is possible 
even for maxi\-mal possible $m^{(3)}_{ee}$ (figs. \ref{lmatrans} b,c)).
In contrast, 
for a lower $\sin^2 2\theta_{\odot}$ the cancellation can be
complete so that no lower bound appears (see fig. \ref{lmatrans} a)).

In the region of $m_1 \simeq 10^{-3}$ eV
all states contribute with comparable portions 
to $\langle m \rangle$, 
thus cancellation is possible and no lower bound exists.

For larger values of $m_1$ the first and the second state give the dominating
contributions to $\langle m \rangle$ and the increase of $m_3$ 
does not influence significantly the total $\langle m \rangle$. 
In this case the mass $\langle m \rangle$ is 
determined by $m_1$ and $\theta_{\odot}$ and a 
larger  $\sin^2 2\theta_{\odot}$ implies a larger  
possible range of $\langle m \rangle$ for a given $m_1$, reflecting the 
uncertainty of unknown Majorana CP-phases.

\begin{figure}[!ht]
\begin{center}
\includegraphics[width=1.0\textwidth]{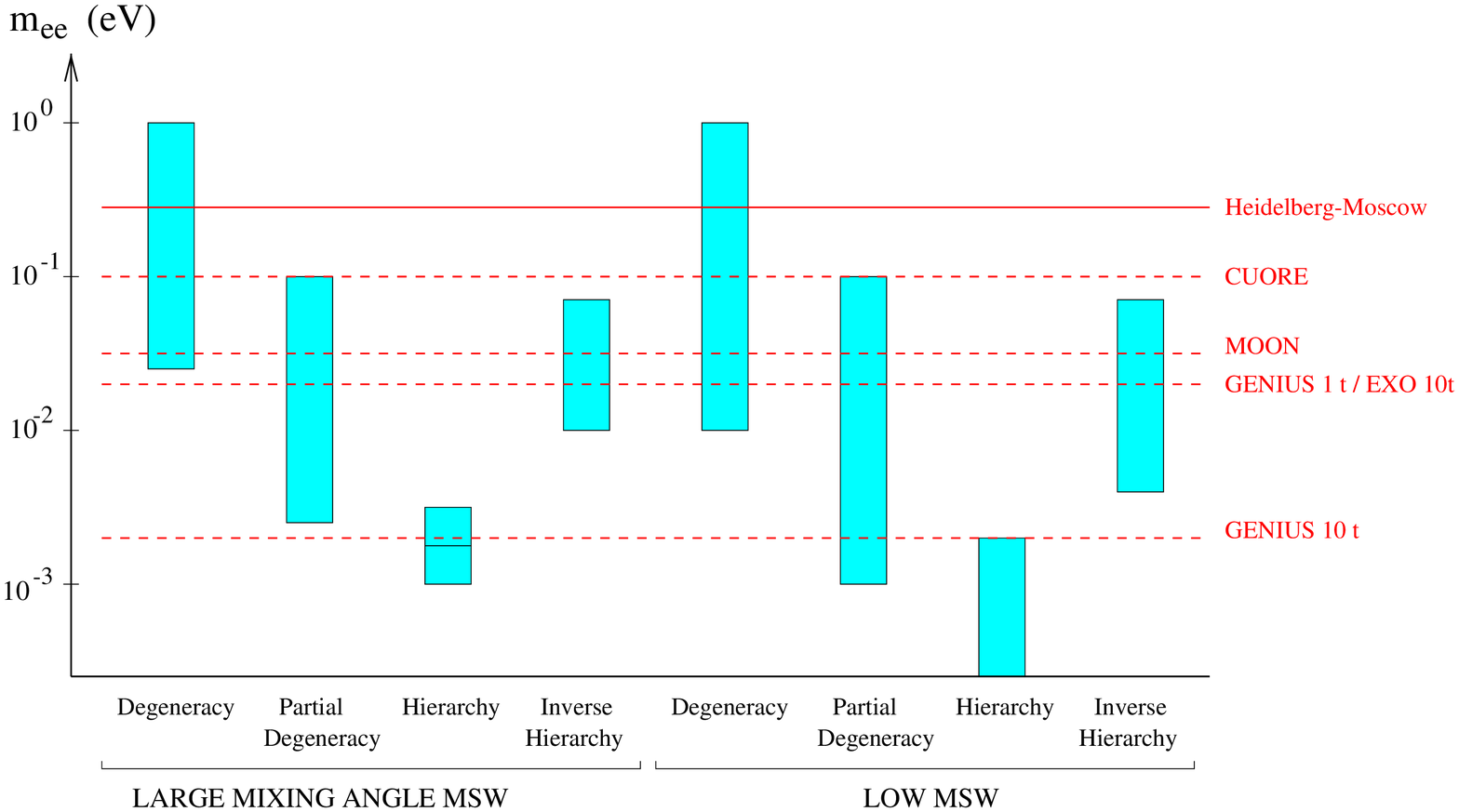}
\end{center}
\caption[]{
Summary of expected values for $\langle m \rangle$ 
in the different schemes discussed
in this paper. The size of the bars corresponds to the uncertainty in
mixing angles and the unknown Majorana CP-phases. 
The expectations are compared with the recent neutrino mass limits 
obtained from the Heidelberg--Moscow \protect{\cite{hdmo}}
experiment as well as the expected 
sensitivities for the CUORE \protect{\cite{cuore}}, MOON 
\protect{\cite{moon}}, EXO \protect{\cite{exo}}
proposals 
and the 1 ton and 10 ton proposal of GENIUS \protect{\cite{gen}}.
\label{cstatessum}}
\end{figure}

\section{Conclusions}
Neutrinoless double beta decay and neutrino oscillations provide 
complementary pieces to the solution of the neutrino mass puzzle.
Correlations of the oscillation parameters and the effective neutrino
Majorana mass $\langle m \rangle$ have been discussed in various scenarios
favored by recent neutrino oscillation data. The new Super-Kamiokande
data presented at the {\it Neutrino2000} conference improve the
prospects of a positive signal in double beta decay. 
Already now an upper bound for the absolute neutrino mass scale of 
$m_{1,2,3}<3$~eV (LOW-QVO) or $m_{1,2,3}<1$~eV (LMA) has been obtained,
being competitive with the recent tritium decay bound \cite{trit}.
A summary of future perspectives is given in
fig. \ref{cstatessum}, where the size of the bars corresponds to the 
uncertainty in mixing angles and the unknown Majorana CP-phases. 
As is obvious from the figure, future double beta
decay projects may be able to test all scenarios but the hierarchical
spectrum with solar neutrino QVO-LOW  solution. One should keep in mind here, 
that the QVO-LOW solution is disfavored in an analysis of the supernova 1987A
\cite{SN87a}.
Depending on the value of 
$\langle m \rangle$ obtained in the future, the follwing conclusions 
can be drawn. 

\begin{itemize}

\item
For $\langle m \rangle > 0.1$~eV the neutrino mass spectrum is degenerate.
An allowed region for the absolute mass scale in the neutrino sector can be 
obtained. Its size depends on the magnitude of mixing of the solar 
neutrinos. If the mixing is large, the uncertainty can be up to a factor 
of 10, if the mixing is small, it will be less than a factor of two. 
For the MSW bestfit it will be about a factor of three. A crucial 
contribution may come from KAMLAND, which has been estimated to
fix $\sin^2 2 \theta_{\odot}$ within $\pm 0.1$ with three years of accumulated 
data \cite{kam}.

\item
For $\langle m \rangle \simeq 0.01 - 0.1$~eV the neutrino mass 
spectrum can be degenerate, partial degenerate or inverse hierarchical.
Again an allowed region for the absolute mass scale can be fixed, provided
the character of hierarchy (direct/inverse) can be established from
neutrino oscillations in matter. A recent analysis comes to the conclusion, 
that the inverse hierarchy is disfavored already for not too large values of
$U_{e3}$.

\item
For $\langle m \rangle \simeq 0.001 - 0.01$~eV the neutrino mass 
spectrum can be partial degenerate or inverse hierarchical.
The conclusions above remain valid.

\item
For $\langle m \rangle < 0.001$~eV the spectrum is hierarchical. 

\end{itemize}

In view of this potential the realization of future double beta decay 
projects is highly desirable. We are entering an exciting decade.

\section{Acknowledgement}
We thank T.J. Weiler for fruitful collaborations this 
review in part is based on. H.P. was supported in part by the
DOE grant no.\ DE-FG05-85ER40226.

\end{document}